\documentclass[a4paper,12pt]{article}
\usepackage{xcolor}
\usepackage{authblk}
\usepackage{graphicx}
\usepackage{epstopdf, epsfig}
\usepackage{listings}
\usepackage{bm}
\usepackage{fancyhdr}
\usepackage{amsmath}
\makeatletter
\def\maketag@@@#1{\hbox{\m@th\normalfont\normalsize#1}}
\makeatother
\usepackage{amsfonts}
\usepackage{amssymb}
\usepackage{amsthm}
\usepackage{mathtools}
\usepackage{caption}
\usepackage[latin1,utf8]{inputenc}
\usepackage{array}
\usepackage{geometry}
\usepackage[english]{nomencl}
\usepackage{comment}

\usepackage{hyperref}

\usepackage{doi}

\usepackage{filecontents}

\title{Similarities between the speed limit in relativity and the sound barrier in inviscid fluid flows} %potential flow}

\author[1]{F. Salmon \thanks{Corresponding Author: Fabien.Salmon@bordeaux-inp.fr}} %\firstname{Fabien} \lastname{Salmon}}
\affil[1]{\small Archéosciences Bordeaux UMR 6034, CNRS, University Bordeaux Montaigne, France}
\date{}
\normalsize
\providecommand{\keywords}[1]{\textbf{\textit{Keywords --}} \small\textit{#1}}

\renewcommand\nomgroup[1]{
	\ifthenelse{\equal{#1}{B}}{
		}{\item[]}{%\textbf{Symboles}]}{
	\ifthenelse{\equal{#1}{G}}{\vspace{0.25cm}
		\item[\textbf{Greek symbols}]}{
	\ifthenelse{\equal{#1}{N}}{\vspace{0.25cm}
		\item[\textbf{Nombres sans dimension}]}{
	\ifthenelse{\equal{#1}{R}}{\vspace{0.25cm}
		\item[\textbf{Exposants}]}{
	\ifthenelse{\equal{#1}{Q}}{\vspace{0.25cm}
		\item[\textbf{Indices}]}{
	\ifthenelse{\equal{#1}{V}}{\vspace{0.25cm}
		\item[\textbf{Acronyms}]}{
	\ifthenelse{\equal{#1}{P}}{\vspace{0.25cm}
		\item[\textbf{Physical constants}]}{
	\ifthenelse{\equal{#1}{X}}{\vspace{0.25cm}
		\item[\textbf{Others}]}{
	{}}}}}}}}}}

\newcommand{\nomunit}[1]{
	\renewcommand{\nomentryend}{\hspace*{\fill}#1}}
\makenomenclature

\counterwithin*{equation}{section}

\begin{document}

% Use the \maketitle command after the abstract
\maketitle

\begin{abstract} %an inviscid fluid and with no vorticity present in the flow.
The Prandtl–Glauert–Lorentz transformation establishes a deep mathematical correspondence between compressible potential flow theory and special relativity. Despite being implicit in the aerodynamic literature since the early twentieth century, this analogy remains largely unrecognised across the two communities and has never been surveyed in a unified manner. This review presents, in a self-contained and pedagogical framework accessible to both communities, the historical, mathematical, and applied dimensions of this correspondence. After tracing the parallel developments that gave rise to compressible aerodynamic theory and special relativity, we establish the formal equivalence between the Lorentz transformation and the Prandtl–Glauert coordinate change, and reinterpret relativistic concepts such as time dilation, the Doppler effect, and the Lorentz factor within the acoustic framework. Thanks to the added mass concept, we finally extend this analogy and show that the whole mathematical framework of special relativity is exactly mirrored by the linearised compressible potential flow theory.
\end{abstract}


\printnomenclature[2.5cm]

\nomenclature[p]{$c$}{Speed of light\nomunit{$\sim 3\times 10^8$ m$\cdot$s\textsuperscript{-1}}}
\nomenclature[p]{$e$}{Elementary charge\nomunit{$\sim 1.6\times 10^{-19}$ A$\cdot$s\textsuperscript{2}}}
\nomenclature[p]{$k_B$}{Boltzmann constant\nomunit{$\sim 1.38\times 10^{-23}$ kg$\cdot$m\textsuperscript{2}$\cdot$s\textsuperscript{-2}$\cdot$K\textsuperscript{-1}}}
\nomenclature[p]{$\varepsilon_0$}{Vacuum permittivity\nomunit{$\sim 8.85\times 10^{-12}$ A\textsuperscript{2}$\cdot$s\textsuperscript{4}$\cdot$kg\textsuperscript{-1}$\cdot$m\textsuperscript{-3}}}
\nomenclature[b]{$c_s$}{Speed of sound\nomunit{m$\cdot$s\textsuperscript{-1}}}
\nomenclature[b]{$\bm{v}$}{Body or particle velocity\nomunit{m$\cdot$s\textsuperscript{-1}}}
\nomenclature[b]{$\mathbf{u}$}{Fluid velocity\nomunit{m$\cdot$s\textsuperscript{-1}}}
\nomenclature[b]{$M$}{Mach number $\frac{v}{c_s}$}
\nomenclature[g]{$\beta$}{Relativistic velocity ratio $\frac{v}{c}$}
\nomenclature[g]{$\phi$}{Velocity potential\nomunit{m\textsuperscript{2}$\cdot$s\textsuperscript{-1}}}
\nomenclature[g]{$\gamma$}{Lorentz factor}
\nomenclature[g]{$\gamma_s$}{Prandtl-Glauert factor}
\nomenclature[b]{$\mathbf{p}$}{Relativistic momentum\nomunit{kg$\cdot$m$\cdot$s\textsuperscript{-1}}}
\nomenclature[b]{$\mathbf{F}$}{External forces \nomunit{N}}
\nomenclature[b]{$m$}{Body or particle mass\nomunit{kg}}
\nomenclature[b]{$m_a$}{Virtual added mass\nomunit{kg}}
\nomenclature[b]{$E$}{Energy\nomunit{J}}
\nomenclature[b]{$K$}{Kinetic energy\nomunit{J}}
\nomenclature[b]{$T$}{Period\nomunit{s}}
\nomenclature[b]{$f$}{Frequency\nomunit{s\textsuperscript{-1}}}
\nomenclature[b]{$k$}{Angular wavenumber\nomunit{rad$\cdot$m\textsuperscript{-1}}}
\nomenclature[g]{$\lambda$}{Wavelength\nomunit{m}}
\nomenclature[g]{$\rho$}{Fluid density\nomunit{kg$\cdot$m\textsuperscript{-3}}}
\nomenclature[b]{$P$}{Fluid pressure\nomunit{Pa}}
\nomenclature[b]{$P_0$}{Far-field pressure\nomunit{Pa}}
\nomenclature[b]{$p$}{Fluid pressure variation\nomunit{Pa}}
\nomenclature[g]{$\rho_0$}{Far-field density\nomunit{kg$\cdot$m\textsuperscript{-3}}}
\nomenclature[q]{comp}{Compressible fluid}
\nomenclature[q]{incomp}{Incompressible fluid}
\nomenclature[q]{cr}{Critical value corresponding to a local Mach number of unity}
\nomenclature[b]{$C_p$}{Pressure coefficient}
\nomenclature[g]{$\kappa$}{Heat capacity ratio}
\nomenclature[g]{$\omega$}{Angular frequency\nomunit{rad$\cdot$s\textsuperscript{-1}}}
\nomenclature[g]{$\rho_c$}{Critical density of the Universe\nomunit{kg$\cdot$m\textsuperscript{-3}}}
\nomenclature[b]{$T_{CMB}$}{Temperature of the Cosmic Microwave Background \nomunit{K}}

\section{Introduction}
The history of theoretical physics and fluid mechanics offers numerous examples where seemingly unrelated fields share deep mathematical structures (electromagnetism, sonic black holes, \textit{etc}). Among the most striking, and perhaps least appreciated, is the formal analogy between the theory of compressible potential flows and the kinematics of special relativity. The Prandtl-Glauert transformation, involved in compressible aerodynamics, and the Lorentz transformation of special relativity are not merely superficially similar: they share the same mathematical skeleton, with the speed of sound playing the role of the speed of light, and the Mach number, defined as the ratio of the fluid velocity to the speed of sound $M=\frac{v}{c_s}$, that of the relativistic velocity ratio $\beta=\frac{v}{c}$. This correspondence could just be a simple curiosity or be a real physical link, even though both disciplines have developed largely independently of one another. \newline

The origins of this analogy trace back to the nineteenth century. The wave nature of light, as described by electromagnetism and Maxwell's equations, originally led physicists to postulate the existence of a propagation medium known as the luminiferous aether. In the 1880s, although this concept was widely accepted, A. Michelson and E. Morley \cite{Michelson} conducted a series of experiments that did not give the expected results. To account for them, special relativity was developed by several physicists. Mainly formulated by Einstein in 1905 on the basis of two deceptively simple postulates, the theory replaces the Galilean framework of classical mechanics with a geometry of space-time governed by the Lorentz transformation. To date, special relativity has successfully passed all experimental tests \cite{Test_RR, Test_RR2, Test_RR3} and the Lorentz transformation has been experimentally validated to an extraordinary degree of precision, including in regimes extremely close to the speed of light. In particle accelerators such as the Large Hadron Collider (LHC) at CERN, protons are accelerated to Lorentz factors of $\gamma \approx 7500$, corresponding to velocities of $v \approx 0.999999991 c$, about 3 m$\cdot$s$^{-1}$ below the speed of light. In parallel, the rapid expansion of aviation in the 1920s and 1930s drove the development of compressible aerodynamic theory. To account for the compressible effects, Prandtl and then Glauert developed their correction applicable to subsonic compressible flows \cite{Glauert, Prandtl}. The mathematical equivalence between the two frameworks was long implicit in the literature, even if it was already noted by Küssner \cite{Kussner2}, before being fully acknowledged and sometimes renamed the "Prandtl–Glauert–Lorentz transformation". To the best of the author's knowledge, the term was first introduced in 2019 in \cite{n1}. \newline

The purpose of this review is to document and consolidate this analogy in a self-contained and pedagogical manner, tracing its historical roots, establishing its mathematical foundations, and surveying the applications of this transformation across theoretical, numerical, and engineering contexts. The scope intentionally spans both sides of the correspondence: the relativistic framework is presented so that readers from the fluid mechanics and aeroacoustics communities may engage with the analogy on familiar terms. Conversely, the compressible flow framework is detailed to provide a pedagogical content to a broader audience in theoretical physics. However, this review will not address potential physical interpretations of this mathematical correspondence. \newline

The review is organised as follows. Section 2 provides a historical account of the parallel developments in fluid mechanics and theoretical physics that gave rise to the two frameworks, from the Navier–Stokes equations and the genesis of compressible aerodynamic theory to the formulation of special relativity and the early experimental debates surrounding it. It ends with the contemporary use of the Prandtl-Glauert-Lorentz transformation. Section 3 first presents the theoretical framework of special relativity, covering the Lorentz transformations, the relativistic Doppler effect, relativistic dynamics, and the mass-energy equivalence. Then, the derivation of the compressible potential flow theory is detailed.
Based on the added-mass concept, Section 4 presents an extension of this analogy to a strict equivalence between the mathematical frameworks of linearised compressible potential flows and special relativity, even for dynamics.

\section{Historical overview}
This section does not aim to provide an exhaustive account of the history of fluid mechanics and theoretical physics, but rather to trace the development of the key ideas relevant to this review, and to contextualise each theoretical advance within the state of knowledge of its time.

\subsection{The early history of fluid mechanics}
The mathematical foundations of fluid mechanics were laid in the eighteenth century, when the application of differential calculus to physical problems transformed the field from qualitative observation into rigorous theory. In 1738, Daniel Bernoulli formulated the first quantitative laws governing inviscid fluid motion by invoking the principle of conservation of mechanical energy, establishing a direct relationship between flow velocity and pressure that remains a cornerstone of the discipline. A decade later, Jean le Rond d'Alembert \cite{Alembert} introduced the concepts of internal pressure, velocity fields, and partial derivatives applied to fluid continua, providing the mathematical language upon which all subsequent developments would build. These foundations were further consolidated by Leonhard Euler \cite{Euler}, who derived the equations of motion for ideal fluids in their modern form, demonstrating that differential equations could describe fluid dynamics without recourse to restrictive physical hypotheses. \newline

A critical simplifying hypothesis emerged naturally from these equations: if, in addition to being inviscid and incompressible, the flow is assumed irrotational, the curl of the velocity field vanishes and the entire velocity field can be derived from a single scalar function, the velocity potential $\phi$, through $\mathbf{u} = \boldsymbol{\nabla} \phi$. This concept was introduced by Lagrange in 1781 in \cite{Lagrange}, as a means to reduce the Euler equations for incompressible fluids to a more manageable form involving scalar potentials. Under these three assumptions (inviscid, incompressible, and irrotational), the stationary governing equation collapses from the full nonlinear Euler system to the linear Laplace equation, $\nabla^2 \phi = 0$, a simplification that permits closed-form analytical solutions.\newline

The nineteenth century brought two critical extensions to Euler's framework.  First, was the incorporation of viscosity into the governing equations. Navier introduced internal friction as an additional term in the equation of fluid motion in 1822 \cite{Navier}. The same equation was subsequently re-derived by Cauchy, Poisson, and Saint-Venant, each using different molecular or continuum assumptions and each largely ignoring or criticising the work of predecessors. The definitive formulation came from Stokes (1845) \cite{Stokes}, who abandoned molecular hypotheses entirely in favour of a continuum mechanics framework, thereby giving the equations a more general and enduring continuum-mechanical foundation. A more detailed historical description of the genesis of the Navier-Stokes equations is given by O. Darrigol \cite{Darrigol}. Second, the theoretical treatment of compressible flows developed progressively throughout the nineteenth century, driven primarily by interest in acoustics and wave propagation rather than by engineering applications. The pioneering thermodynamic work of Laplace (1816) supplied the closure equation that the compressible Euler system had lacked since its inception. Building on this foundation, Riemann \cite{Riemann} (1860) provided the first rigorous analysis of nonlinear wave propagation in compressible inviscid flows and established the mathematical existence of shock waves. The jump conditions relating flow quantities across such discontinuities were subsequently derived on thermodynamic grounds by Rankine \cite{Rankine} in 1870 and completed by Hugoniot (1887–1889), yielding the Rankine–Hugoniot relations that remain the cornerstone of shock wave theory. It should be noted, however, that theoretical activity in compressible flows remained comparatively limited throughout the nineteenth century, as the absence of high-speed engineering applications such as supersonic aerodynamics and turbomachinery provided little practical incentive for their systematic investigation.

\subsection{Special relativity birth}
This section provides a non-exhaustive account of the genesis of special relativity. Readers seeking a comprehensive historical treatment are referred to \cite{Miller}. \newline

The origin of special relativity cannot be understood in isolation from the nineteenth-century crisis in electromagnetism and optics. The prevailing mechanical worldview held that light, like sound, required a propagating medium, the so-called luminiferous aether, whose existence was regarded as a physical necessity. Following the wave theory of light consolidated by Thomas Young (1804) and Augustin-Jean Fresnel (1816), this hypothetical medium was modelled as a rigid elastic solid capable of sustaining transverse oscillations. The first deep theoretical challenge to this picture came from James Clerk Maxwell, whose 1864 unification of electricity, magnetism, and optics into a single set of field equations implied that light was an electromagnetic wave propagating at a universal speed $c$ \cite{Maxwell}. In 1887, Heinrich Hertz demonstrated the existence of electromagnetic waves, leading to the acceptation of Maxwell's theory by the physicist community. While this theory was mathematically complete, it offered no satisfactory mechanical description of the aether, and its treatment of moving bodies remained problematic. \newline

The experimental situation became increasingly acute in the end of the nineteenth century. Two key experiments, whose results were mutually contradictory within the aether framework, dominated theoretical discussion. On one hand, the Fizeau experiment (1851) measured the speed of light in moving water and appeared to confirm Fresnel's partial-dragging coefficient, lending support to the stationary aether hypothesis. On the other hand, the celebrated Michelson–Morley experiment (1887), designed to detect the motion of the Earth through the aether by measuring differences in the speed of light along perpendicular directions, returned a null result with unprecedented precision \cite{Michelson}. Together, these two outcomes were irreconcilable within any single consistent aether model: Fresnel's stationary aether explained the Fizeau result but not the Michelson–Morley null result, while a fully dragged aether, as proposed by Stokes and later Hertz, accommodated the latter but was incompatible with the observed aberration of starlight. \newline

Theoretical responses to this crisis were numerous and increasingly sophisticated, yet each fell short of a complete resolution. Woldemar Voigt (1887) derived, in the context of wave propagation in an incompressible elastic medium, a set of coordinate transformations that left the wave equation invariant and implicitly contained what would later be identified as the Lorentz factor $\gamma=\left(1-\frac{v^2}{c^2}\right)^{-\frac{1}{2}}$ and a "local time" variable. His work was, however, entirely ignored by his contemporaries. George FitzGerald (1889) and Hendrik Lorentz (1892) independently proposed that material bodies contract in their direction of motion through the aether as a way of explaining the null result of the Michelson–Morley experiment, building on Heaviside's earlier analysis of the deformation of electrostatic fields in motion. Building on these ideas, Lorentz developed a systematic aether theory \cite{Lorentz} in which he derived, in 1892 and 1895, the full set of coordinate transformations relating the aether rest frame to a moving frame. The key ingredient was a modified time variable, the "local time" $t' = t - \frac{vx}{c^2}$, which encodes the fact that, for a moving observer, clocks at different positions along the direction of motion appear to be offset relative to one another, departing from the universal time $t$ of the stationary aether. Crucially, however, Lorentz regarded local time as a purely auxiliary mathematical construct devoid of physical meaning: it was a calculational tool, not a statement about what clocks actually measure. \newline

Henri Poincaré came considerably closer to the final theory. From 1900 onwards he recognised that "local time" is precisely what moving clocks indicate, identified the group structure of the Lorentz transformations, introduced the relativity principle as a general law of nature, and by 1905 had independently derived the transformation equations in their complete form \cite{Poincare}. Yet Poincaré retained the aether as a conceptual anchor, continuing to distinguish between the "true" coordinates of an observer at rest in the aether and the "apparent" coordinates of a moving observer. Einstein \cite{Einstein} discarded the aether entirely and built the entire theory on two postulates: the principle of relativity (the laws of physics take the same form in all inertial frames) and the principle of the constancy of the speed of light (light propagates at speed $c$ in vacuum independently of the motion of the source or observer). From these two postulates alone, without any reference to the constitution of matter or to electromagnetic models, Einstein derived the Lorentz transformations, time dilation, length contraction, the relativity of simultaneity, and the velocity addition formula. The elimination of the aether removed the need to distinguish true from apparent coordinates: all inertial frames are physically equivalent, and no preferred rest frame exists. This conceptual economy, absent from all earlier formulations, is what distinguishes Einstein's contribution from those of Lorentz and Poincaré, despite the mathematical equivalence of many results. \newline

By approximately 1911, special relativity had gained acceptance among the majority of theoretical physicists, and the aether hypothesis was progressively abandoned \cite{Whittaker}. Only a minority of theoretical physicists such as Abraham, Lorentz, or Poincaré still believed in the existence of an aether. Poincaré described light as "luminous vibrations of the aether" as late as 1912, and argued in his posthumously published \textit{Dernières pensées} (1913) that physicists who adopted the new relativistic conventions did so merely because they found them more convenient, not because they were compelled to by physical reality. For Poincaré, special relativity was an elegant mathematical convention rather than a definitive statement about the physical world. Lorentz, until his death in 1928, maintained that a physically real privileged rest frame, the aether, must exist, even if the laws of physics conspire to render it experimentally inaccessible. He accepted the mathematical framework of special relativity while rejecting its philosophical conclusion that absolute rest is a meaningless notion. Lorentz's aether theory and Einstein's special relativity are empirically equivalent: they produce identical predictions for every conceivable experiment, the sole difference being the metaphysical postulate of an undetectable absolute rest frame. No experiment was thought capable of adjudicating between the two theories. All arguments historically advanced in favour of Einstein's formulation over Lorentz's rest on non-empirical criteria such as conceptual economy, parsimony, and explanatory power rather than on experimental evidence \cite{Acuna}. The triumph of special relativity was thus influenced, in part, by philosophical considerations.

\subsection{Early aviation and the myth of the sound barrier}
The theoretical foundations of aerodynamics developed in close interaction with the rapid expansion of aviation engineering during the first half of the twentieth century. The initial framework rested entirely on incompressible potential flow theory. Following the work of Helmholtz on vortex dynamics (1858), the concept of circulation as the physical mechanism for aerodynamic lift was independently proposed by Frederick Lanchester during the 1890s but published in 1907 \cite{Lanchester} and formalised mathematically by Kutta in 1902 \cite{Kutta} and Joukowski in 1906. The latter developed the conformal mapping transformation that converts flow around a circular cylinder into flow around an airfoil shape. Just three years after the Wright brothers' first powered flight, the circulation theory of lift was in place, ready to aid the design and understanding of lifting surfaces. These two-dimensional results were extended to finite wings by the Lanchester–Prandtl lifting-line theory \cite{Prandtl2}, which represented the wing as a bound vortex line along its span, accounting for the trailing vortex sheet shed from the trailing edge and enabling the calculation of induced drag for finite-span wings.\newline

The other transformative contribution of this period came from Prandtl at Göttingen. In 1904, Prandtl delivered his landmark paper \cite{Prandtl1} on the boundary layer at the Third International Mathematics Congress in Heidelberg, in which he introduced the concept that viscous effects are confined to a thin region adjacent to the body surface, described flow separation, and provided the first theoretical explanation of aerodynamic stall. This single insight resolved d'Alembert's paradox by reconciling inviscid outer flow with viscous inner flow, and established the conceptual split between boundary layer theory and potential flow that underlies all subsequent aerodynamic analysis. \newline

As aircraft speeds increased through the 1920s and 1930s, the incompressibility assumption became increasingly untenable. The critical question was how to correct aerodynamic predictions for compressibility effects at high subsonic Mach numbers. In the 1920s, Prandtl and then Glauert developed the first correction rule for accounting for compressibility effects on airfoil lift coefficients at high subsonic speeds. The resulting Prandtl–Glauert transformation, formally published by Glauert in 1928 \cite{Glauert}, is a linearisation of the compressible potential flow equation for thin bodies in subsonic freestreams: it relates the compressible solution to an equivalent incompressible problem by the scaling factor $(1-M^2)^{-\frac{1}{2}}$ where $M$ is the Mach number, i.e., the Lorentz factor based on the speed of sound. Its theoretical development will be detailed in the next section. Prandtl and Glauert therefore predicted a divergence of aerodynamic forces as the freestream Mach number approaches unity. This transformation was extended by Ackeret to supersonic freestreams in 1925 \cite{Ackeret}, providing an analogous linearised theory for supersonic thin-airfoil aerodynamics where the Lorentz factor is replaced with $(M^2-1)^{-\frac{1}{2}}$. The transformation was then further studied by Küssner \cite{Kussner1, Kussner2} and extended to the unsteady case. Recognising the inaccuracy of the Prandtl–Glauert rule at high subsonic Mach numbers, von Kármán and Tsien \cite{Tsien,Karman} proposed a nonlinear correction, improving accuracy in the high-subsonic regime. In 1951, Laitone \cite{Laitone} enhanced this formulation by giving a more complex expression of the pressure coefficient. \newline

The structural limitation of all these approaches, linearised or mildly nonlinear potential theory, is their fundamental breakdown in the transonic regime. When the freestream Mach number lies between roughly 0.8 and 1.2, regions of locally supersonic flow coexist with subsonic regions, shock waves form on the wing surface, and the governing equations become nonlinear in a manner that cannot be linearised. This theoretical void had direct engineering consequences: in the 1930s and early 1940s, pilots and engineers encountered violent buffeting, loss of control, and structural stress as aircraft approached Mach 1, fuelling a widespread belief that the speed of sound represented an impassable physical barrier. The practical demonstration that the sound barrier was not physically impassable but an engineering challenge came on 14 October 1947, when Chuck Yeager reached Mach 1.07 in the Bell X-1. Yeager's achievement demonstrated that the "barrier" was a transitory aerodynamic phenomenon, not a fundamental physical limit.

\subsection{On the Modern Use of the Prandtl–Glauert–Lorentz Transformation}
Despite the great advances in high-performance computing that have made direct numerical approaches increasingly attractive, the Prandtl-Glauert-Lorentz transformation has retained its relevance in the fluid mechanics and aeroacoustics communities, owing to the fundamental analytical and numerical simplifications it affords. \newline

Some authors have worked on the transformations itself, attempting to improve them or testing them. Liu \cite{Liu} proposed a new similarity law to improve the treatment of steady transonic-supersonic flows over thin bodies. In his dissertation, B. J. German \cite{German} developed a Riemannian geometric framework to identify incompressible equivalents of nonlinear subsonic compressible potential flows. The work generalizes classical compressibility transformations such as the Prandtl–Glauert and Kármán–Tsien corrections by introducing nonlinear coordinate and metric mappings that recast the compressible potential equation into an incompressible Laplacian form. Kirkman and Takahashi \cite{Jeffrey} claimed that several historical interpretations of compressibility corrections are mathematically ambiguous or incomplete, and they proposed a reformulation to obtain a more consistent prediction of the critical Mach number. The proposed approach was then assessed through comparisons with panel-method and CFD (Computational Fluid Dynamics) solutions. Even today, the Prandtl-Glauert-Lorentz transformation is still used in some studies. Chandre-Vila et al. \cite{Chandre} developed a simplified methodology to compute the nonlinear aeroelastic effects of high-aspect-ratio wings, where the compressibility effects are assessed with the Prandtl-Glauert transformation. In \cite{Wang}, the authors compared the performance of the Prandtl–Glauert-Lorentz, Karman–Tsien, and Laitone transformations for the computation of the compressibility-corrected pressure coefficient of underwater vehicles. In particular, they were also compared to CFD simulations. Hong et al. \cite{Hong} computed the compressibility effects using this formulation applied to Blended Wing-Body aircraft while comparing the results with benchmarks. The Prandtl-Glauert-Lorentz transformation was also employed for different applications such as the flight in Martian atmosphere \cite{Mars} or nacelle design \cite{Nacelle}. \newline

A key result of the Prandtl-Glauert-Lorentz transformation is the similarity between the classical wave equation and the convected wave equation first noted in the aeroacoustic context by Amiet and Sears \cite{frequency}. Then, Taylor \cite{Taylor} explored the implications of this transformation on wind-tunnel and flight-test corrections. Chapman \cite{Chapman} subsequently derived a parametric family of similarity variables that unifies Doppler factors and Prandtl–Glauert-Lorentz coordinates and establishes a general "fundamental rule" for transferring free-field solutions to the convected case. The geometric structure underlying this transformation was made explicit by Gregory et al. \cite{Gregory1, Gregory2}, who introduced the concept of an acoustic space-time in direct analogy with the Minkowskian space-time of special relativity, and used geometric algebra to provide a coordinate-free formulation with applications to Doppler shifts and Green's function derivations in uniform flow. More recently, the transformation has been employed to unify the stationary and convective formulations of the Ffowcs Williams–Hawkings (FW-H) equation \cite{Du}, yielding compact integral solutions for arbitrary mean flow orientations. \newline

The transformation has also proven particularly valuable for the development of numerical solvers for aeroacoustic problems. Since it maps the convected wave equation onto the standard Helmholtz equation, it allows classical numerical tools such as boundary element methods (BEM), finite element methods (FEM), and perfectly matched layers (PML), to be reused directly in the presence of mean flow, without requiring the development of dedicated convective solvers. Balin et al. \cite{Balin} developed a coupled BEM-FEM strategy in which the Prandtl–Glauert-Lorentz transformation is applied in the far-field uniform-flow region to recover the Helmholtz equation, while the finite element method handles the near-field non-uniform flow region. The boundary conditions to be imposed on rigid bodies in the transformed domain were analysed rigorously by Hu et al. \cite{n1}, who showed that the Zero Energy Flux solid wall condition reduces to the standard zero-normal-velocity condition under the transformation, thereby preserving the simplicity of classical BEM kernels. For absorbing boundary treatments, Barucq et al. \cite{Barucq} designed stable low-order perfectly matched layers based on the Prandtl–Glauert–Lorentz transformation for the convected Helmholtz equation discretised with discontinuous Galerkin methods.

\section{Physical background}
\subsection{Fundamentals of special relativity}
Special relativity rests on two postulates: the laws of physics take the same form in all inertial reference frames, and the speed of light in vacuum $c$ is the same for all inertial observers, regardless of the motion of the source. From these two axioms alone, the entire kinematic and dynamic structure of the theory follows deductively. Only the necessary elements of special relativity are presented below. For an exhaustive pedagogical treatment, see \cite{SpecialRelativity}.

\subsubsection{Lorentz transformation}
Consider two inertial frames $\mathcal{S}$ and $\mathcal{S}'$, where $\mathcal{S}'$ moves at constant velocity $v$ along the $x$-axis relative to $\mathcal{S}$. The Lorentz transformation relates the space-time coordinates ($x,y,z,t$) of an event in $\mathcal{S}$ to its coordinates ($x',y',z',t'$) in $\mathcal{S}'$:

\begin{equation*}
x' = \gamma (x - v t), ~ y' = y,~z' = z,~t' = \gamma\left(t - \frac{v}{c^2} x\right)
\end{equation*}

where $\gamma = \left(1-\beta^2\right)^{-\frac{1}{2}}$ is the Lorentz factor and $\beta = \frac{v}{c}$. This transformation replaces the classical Galilean transformation and reduces to them in the limit $v \ll c$. Two consequences follow immediately: length contraction and time dilation. A rod of rest length $L_0$ oriented along the direction of motion appears contracted to an observer in $\mathcal{S}$ as $L = \frac{L_0}{\gamma}$. A clock at rest in $\mathcal{S}'$ ticks more slowly as measured from $\mathcal{S}$. If the clock measures a proper time interval $\Delta \tau$ between two events occurring at the same location in $\mathcal{S}'$, the coordinate time interval measured in $\mathcal{S}$ is $\Delta t = \gamma \Delta \tau$, so that moving clocks run slow relative to a stationary observer.

\subsubsection{Relativistic dynamics}
The extension of Newton's second law to the relativistic regime requires replacing the classical momentum $m\bm{v}$ by the relativistic momentum $\mathbf{p}=\gamma m\bm{v}$, where $m$
is the rest mass of the particle. The relativistic form of Newton's second law is then

\begin{equation}
\frac{\mathrm{d}\mathbf{p}}{\mathrm{dt}} = \frac{\mathrm{d}\gamma m\bm{v}}{\mathrm{dt}} = \mathbf{F}
\label{EqmotionRR}
\end{equation}

This formulation reduces to Newton's second law when $v \ll c$, for which $\gamma \to 1$. A key consequence is that the effective inertia of a body increases with velocity, so that an infinite force would be required to accelerate a massive body to the speed of light, which is therefore an absolute kinematic barrier for massive particles.

\subsubsection{Mass-Energy equivalence}
The work-energy theorem applied to the relativistic equation of motion yields 
\begin{equation}
\mathbf{F} \cdot \bm{v} = \bm{v}\cdot \frac{\mathrm{d}\gamma m \bm{v}}{\mathrm{d}t} = \gamma m \bm{v}\cdot \frac{\mathrm{d} \bm{v}}{\mathrm{d}t} + m v^2  \frac{\mathrm{d} \gamma}{\mathrm{d}t}
\end{equation}
Since $\frac{\mathrm{d} \gamma}{\mathrm{d}t}=\frac{v}{c^2}\gamma^3\frac{\mathrm{d} v}{\mathrm{d}t} \Rightarrow v\gamma\frac{\mathrm{d} v}{\mathrm{d}t} = \frac{c^2}{\gamma^2} \frac{\mathrm{d} \gamma}{\mathrm{d}t}$, the power can be rewritten $\mathbf{F} \cdot \bm{v} = m c^2 \left(\frac{1}{\gamma^2}+\frac{v^2}{c^2} \right) \frac{\mathrm{d} \gamma}{\mathrm{d}t} = m c^2 \frac{\mathrm{d} \gamma}{\mathrm{d}t}$, which leads to the following expression for the total energy of a particle
\begin{equation}
E = \gamma m c^2
\label{mass-energy}
\end{equation}
Equation \eqref{mass-energy} expresses the equivalence of mass and energy as different manifestations of the same physical quantity.

\subsubsection{Relativistic Doppler Effect}
When a source emitting a transverse electromagnetic wave at frequency $f_0$ moves at velocity $v$ relative to an observer at angle $\theta$, the observed frequency $f$ is shifted by two distinct physical mechanisms. The first is the classical wave compression effect: the source partially "chases" the wavefronts it emits, reducing the wavelength received by the observer. In the rest frame of the source, the period of emission is $T_0=\frac{1}{f_0}$. During one period, the source moves a distance $vT_0$ in the direction of motion, so the spatial separation between successive wavefronts is reduced to $\lambda = (c - v\cos\theta)T_0$ instead of $cT_0$. This classical argument alone would give $f=\frac{c}{\lambda}=\frac{f_0}{1-\beta\cos\theta}$. The second mechanism is purely relativistic: time dilation implies that the proper period $T_0$ measured in the source frame corresponds to a dilated period $T= \gamma T_0$ in the observer's frame. So the frequency measured by the observer is given by the relativistic Doppler formula:
\begin{equation}
f = \frac{f_0}{\gamma \left(1-\beta\cos\theta\right)}
\label{Relativistic_Doppler}
\end{equation}

\subsection{Compressible potential flow and Prandtl-Glauert transformation}
This section presents the mathematical formulation of a particle's displacement in a compressible, inviscid fluid. The flow is assumed to be irrotational, so there exists a velocity potential $\phi$ such that $\mathbf{u} = \boldsymbol{\nabla} \phi$. For simplicity, we consider the particle moving at velocity $v\mathbf{e_x}$. In the particle's reference frame, far from the particle, the fluid then moves at $-v\mathbf{e_x}$. Under these assumptions, the equations governing the fluid are
\begin{equation}
\begin{array}{lc}
\frac{1}{\rho}\frac{\partial \rho}{\partial t} + \boldsymbol{\nabla}\cdot\mathbf{u} + \frac{1}{\rho}\mathbf{u}\cdot \boldsymbol{\nabla} \rho = 0 \\
\frac{\partial \mathbf{u}}{\partial t} + \frac{1}{2}\boldsymbol{\nabla} (\mathbf{u}\cdot\mathbf{u}) = -\frac{\boldsymbol{\nabla} P}{\rho}
\end{array}
\label{eqRRB}
\end{equation}
where $P$ is the fluid pressure and $\rho$ is its density. Introducing $\mathbf{u} = \boldsymbol{\nabla} \phi$ in the momentum equation leads to
\begin{equation}
\frac{\partial \boldsymbol{\nabla} \phi}{\partial t} + \frac{1}{2}\boldsymbol{\nabla} (\boldsymbol{\nabla} \phi\cdot\boldsymbol{\nabla} \phi) = -\frac{\boldsymbol{\nabla} P}{\rho} \label{eqini}
\end{equation}
The left-hand term represents a gradient, which necessitates that the right-hand term must also be a gradient. Consequently, the compressible potential flow hypothesis imposes the condition $\boldsymbol{\nabla} \times \left( \frac{\boldsymbol{\nabla} P}{\rho} \right) = \mathbf{0}$. This yields the relation $\boldsymbol{\nabla} \left(\frac{1}{\rho}\right) \times \boldsymbol{\nabla} P = -\frac{1}{\rho^2} \boldsymbol{\nabla}\rho \times \boldsymbol{\nabla} P = \mathbf{0}$. As a result, $\boldsymbol{\nabla} P $ is proportional to $\boldsymbol{\nabla} \rho$, implying that the fluid is barotropic, i.e., $p=f(\rho)$. Thus, it follows that

\begin{align}
 & \frac{\partial \boldsymbol{\nabla} \phi}{\partial t} + \frac{1}{2}\boldsymbol{\nabla} (\boldsymbol{\nabla} \phi\cdot\boldsymbol{\nabla} \phi) = -\frac{\boldsymbol{\nabla} P}{\rho} = -\boldsymbol{\nabla} \int_{P_0}^P \frac{d\tilde{P}}{\rho} \\
\Rightarrow & \frac{\partial \phi}{\partial t} + \frac{1}{2} (\boldsymbol{\nabla} \phi\cdot\boldsymbol{\nabla} \phi) + \int_{P_0}^P \frac{d\tilde{P}}{\rho} = C(t) \label{eqini2} \\
\Rightarrow &  \frac{\partial^2 \phi}{\partial t^2} + \frac{1}{2} \frac{\partial}{\partial t}\left(\boldsymbol{\nabla} \phi\cdot\boldsymbol{\nabla} \phi\right) + \frac{1}{\rho} \frac{\partial P}{\partial t} = 0
\end{align}
where $C(t)$ is an arbitrary function. As $\phi$ is not uniquely defined, we can set $C(t) = 0$ without loss of generality.

The definition of the speed of sound gives $c_s^2=\frac{dP}{d\rho}$. Given that the fluid is barotropic, assuming constant the speed of sound yields $c_s^2 \frac{\partial \rho}{\partial t} = \frac{\partial P}{\partial t}$. The momentum equation thus becomes

\begin{equation}
\frac{\partial^2 \phi}{\partial t^2} + \frac{1}{2} \frac{\partial}{\partial t}\left(\boldsymbol{\nabla} \phi\cdot\boldsymbol{\nabla} \phi\right) + \frac{c_s^2}{\rho} \frac{\partial \rho}{\partial t} = 0 \Rightarrow \frac{1}{\rho} \frac{\partial \rho}{\partial t} = -\frac{1}{c_s^2} \frac{\partial}{\partial t} \left(\frac{\partial \phi}{\partial t} + \frac{1}{2} \left(\boldsymbol{\nabla} \phi\cdot\boldsymbol{\nabla} \phi\right)\right)
\end{equation}

Given equation \eqref{eqini}, we have
\begin{equation}
\frac{1}{\rho}\mathbf{u}\cdot \boldsymbol{\nabla} \rho = \frac{1}{\rho c_s^2}\mathbf{u}\cdot \boldsymbol{\nabla} P = -\frac{1}{c_s^2}\boldsymbol{\nabla} \phi\cdot \boldsymbol{\nabla}\left(\frac{\partial \phi}{\partial t} + \frac{1}{2} \left(\boldsymbol{\nabla} \phi\cdot\boldsymbol{\nabla} \phi\right)\right)
\end{equation}

Therefore, the continuity equation of \eqref{eqRRB} becomes

\begin{align}
 & \nabla^2 \phi -\frac{1}{c_s^2} \frac{\partial}{\partial t} \left(\frac{\partial \phi}{\partial t} + \frac{1}{2} \left(\boldsymbol{\nabla} \phi\cdot\boldsymbol{\nabla} \phi\right)\right) -\frac{1}{c_s^2}\boldsymbol{\nabla} \phi\cdot \boldsymbol{\nabla}\left(\frac{\partial \phi}{\partial t} + \frac{1}{2} \left(\boldsymbol{\nabla} \phi\cdot\boldsymbol{\nabla} \phi\right)\right) = 0 \\
\Rightarrow & \nabla^2 \phi -\frac{1}{c_s^2} \left( \frac{\partial^2 \phi}{\partial t^2} + \frac{\partial}{\partial t}\left(\boldsymbol{\nabla} \phi\cdot\boldsymbol{\nabla} \phi\right) + \frac{1}{2}\boldsymbol{\nabla} \phi\cdot \boldsymbol{\nabla}\left(\boldsymbol{\nabla} \phi\cdot\boldsymbol{\nabla} \phi\right)\right) = 0 
\end{align}

By noting $x_i$ with $i \in [0,2]$ representing coordinates $x$, $y$ and $z$ respectively, this relation becomes
\begin{equation}
\sum_{i=0}^2 \frac{\partial^2 \phi}{\partial x_i^2} - \frac{1}{c_s^2}\left( \frac{\partial^2 \phi}{\partial t^2} + \frac{\partial}{\partial t}\left( \sum_{i=0}^2 \left(\frac{\partial \phi}{\partial x_i}\right)^2\right) + \sum_{i=0}^2 \sum_{k=0}^2 \frac{\partial \phi}{\partial x_i} \frac{\partial \phi}{\partial x_k} \frac{\partial^2 \phi}{\partial x_i \partial x_k} \right) = 0
\label{Add}
\end{equation}

Introducing the cycling index $j \equiv i+1 \pmod 3$, the double sum can be rewritten as
\begin{equation}
\sum_{i=0}^2 \sum_{k=0}^2 \frac{\partial \phi}{\partial x_i} \frac{\partial \phi}{\partial x_k} \frac{\partial^2 \phi}{\partial x_i \partial x_k} = \sum_{i=0}^2 \left( \frac{\partial \phi}{\partial x_i} \right )^2 \frac{\partial^2 \phi}{\partial x_i^2} + 2 \frac{\partial \phi}{\partial x_i} \frac{\partial \phi}{\partial x_j} \frac{\partial^2 \phi}{\partial x_i \partial x_j}
\end{equation}

Then, \eqref{Add} becomes

\begin{equation}
\frac{1}{c_s^2}\frac{\partial^2 \phi}{\partial t^2} = \sum_{i=0}^2 \left[\left(1-\left(\frac{\frac{\partial \phi}{\partial x_i}}{c_s}\right)^2\right) \frac{\partial^2 \phi}{\partial x_i^2} - \frac{2}{c_s^2} \left(\frac{\partial \phi}{\partial x_i} \frac{\partial^2 \phi}{\partial x_i \partial t} + \frac{\partial \phi}{\partial x_i} \frac{\partial \phi}{\partial x_j} \frac{\partial^2 \phi}{\partial x_i \partial x_j} \right)\right]
\label{eq_comp0}
\end{equation}

Under the assumption of small perturbations in the flow, third order terms are neglected. By considering the Mach number in each direction $M_i = \frac{u_i}{c_s} = \frac{\frac{\partial \phi}{\partial x_i}}{c_s}$, equation \eqref{eq_comp0} becomes

\begin{equation}
\frac{1}{c_s^2}\frac{\partial^2 \phi}{\partial t^2} - \sum_{i=0}^2 \left(1-M_i^2\right) \frac{\partial^2 \phi}{\partial x_i^2} + \frac{2}{c_s} \sum_{i=0}^2 M_i \frac{\partial^2 \phi}{\partial x_i \partial t} = 0
\label{eq_comp2}
\end{equation}

Finally, as the fluid velocity is assumed to be mainly along one direction, $\mathbf{u} = (-v + u_x) \mathbf{e_x} + u_y \mathbf{e_y} + u_z \mathbf{e_z}$ where $u_x$, $u_y$ and $u_z$ are small velocity disturbances of the flow far smaller than $v$. The second-order terms in $u_{x_i}$ can also be neglected in equation \eqref{eq_comp2}. The directional Mach number are therefore given by $M_x \sim -\frac{v}{c_s}$ and $M_y\sim M_z\sim 0$. By noting $M=\frac{v}{c_s}$, the sign of the last term must accordingly be modified:

\begin{equation}
\frac{1}{c_s^2}\frac{\partial^2 \phi}{\partial t^2} - \left(1-M^2\right) \frac{\partial^2 \phi}{\partial x^2} - \frac{\partial^2 \phi}{\partial y^2} - \frac{\partial^2 \phi}{\partial z^2} - \frac{2}{c_s} M \frac{\partial^2 \phi}{\partial x \partial t} = 0
\label{eq_comp}
\end{equation}

The corresponding equation in the incompressible limit ($c_s\longrightarrow \infty$) reduces to simply:
\begin{equation}
\frac{1}{c_s^2}\frac{\partial^2 \phi}{\partial t^2} - \frac{\partial^2 \phi}{\partial x^2} - \frac{\partial^2 \phi}{\partial y^2} - \frac{\partial^2 \phi}{\partial z^2} = 0
\label{eq_inc}
\end{equation}

This linearised framework has been widely adopted in the literature, with applications ranging from theoretical sound propagation \cite{Mohring} and the relationship between vorticity and sound \cite{Mohring2} to noise generation by solid bodies \cite{Obermeier}.\newline

The compressible equation \eqref{eq_comp} can be expressed in the form of the incompressible equation \eqref{eq_inc} through a change of coordinates. To do so, we first apply a Galilean transformation to move to the reference frame in which the fluid is at rest:
\begin{equation*}
\tilde{x} = x + v t, ~ \tilde{y} = y,~ \tilde{z} = z,~\tilde{t} = t
\end{equation*}
Next, a Lorentz transformation is applied to the coordinates in the fluid's rest frame:
\begin{equation*}
x' = \gamma_s (\tilde{x} - v \tilde{t}), ~ y' = \tilde{y},~z' = \tilde{z},~t' = \gamma_s\left(\tilde{t} - \frac{v}{c_s^2} \tilde{x}\right)
\end{equation*}
where $\gamma_s = \left(1-\frac{v^2}{c_s^2}\right)^{-\frac{1}{2}}$. The compressible equation \eqref{eq_comp} becomes

\begin{equation}
\frac{1}{c_s^2}\frac{\partial^2 \phi}{\partial t'^2} - \frac{\partial^2 \phi}{\partial x'^2} - \frac{\partial^2 \phi}{\partial y'^2} - \frac{\partial^2 \phi}{\partial z'^2} = 0
\end{equation}

The Prandtl–Glauert transformation \cite{Prandtl,Glauert,Kussner1,Kussner2} maps the linearised compressible potential flow equation onto their incompressible counterpart, thereby allowing compressible aerodynamic characteristics to be deduced directly from incompressible solutions.

\subsection{Speed barrier}
Equation \eqref{eqini2} reads
\begin{equation}
\frac{\partial \phi}{\partial t} + \frac{1}{2} (\boldsymbol{\nabla} \phi\cdot\boldsymbol{\nabla} \phi) + \int_{}^p \frac{d\tilde{P}}{\rho}=0
\label{eqini3}
\end{equation}
The pressure variation is defined as $p=P-P_0$. Let us compute the first-order pressure variation. Keeping only the first-order terms $\frac{1}{2}\boldsymbol{\nabla} \phi\cdot\boldsymbol{\nabla} \phi \sim -v\frac{\partial \phi}{\partial x}$. Given the fluid is barotropic, the Taylor expansion of $\frac{1}{\rho}$ around $P_0$ is
\begin{equation}
\frac{1}{\rho(P)} = \frac{1}{\rho(P_0)} + \left. \frac{\mathrm{d} 1/\rho}{\mathrm{d} P}\right|_{P_0} p + \mathcal{O}(p^2) = \frac{1}{\rho(P_0)} - \frac{1}{\rho(P_0)^2} \left. \frac{\mathrm{d} \rho}{\mathrm{d} P}\right|_{P_0} p + \mathcal{O}(p^2) = \frac{1}{\rho(P_0)} - \frac{p}{c_s^2\rho(P_0)^2} + \mathcal{O}(p^2)
\end{equation}
By noting $\rho_0 = \rho(P_0)$, 
\begin{equation}
\int_{}^p \frac{d\tilde{P}}{\rho} = \frac{p}{\rho_0} + \mathcal{O}(p^2)
\end{equation}
Therefore, retaining only first-order terms, equation \eqref{eqini3} reduces to
\begin{equation}
\frac{\partial \phi}{\partial t} -v\frac{\partial \phi}{\partial x} + \frac{p}{\rho_0} \sim 0
\end{equation}
The first-order pressure perturbation in a compressible potential flow is therefore $p=\rho_0 \left(v\frac{\partial \phi}{\partial x}- \frac{\partial \phi}{\partial t}\right)$. This expression is formally identical to the linearised Bernoulli equation for incompressible potential flow. \newline

To establish the link between the compressible and incompressible solutions, we now show that the temporal terms can be neglected under appropriate conditions, reducing the compressible problem to an incompressible one. In the compressible case, the velocity potential satisfies the wave equation. The temporal term $\frac{1}{c_s^2}\frac{\partial^2 \phi}{\partial t^2}$ can be neglected relative to the spatial terms when $S^2=\left(\frac{La}{c_s v}\right)^2 \ll 1$, where $L$ is a characteristic length of the body and $a$ its acceleration. This condition states that the distance travelled by an acoustic wave during the characteristic time of velocity variation $\frac{v}{a}$ must be large compared to the body size. This quasi-stationary acoustic condition typically applies to small bodies, high wave speeds and high velocities. Under this assumption, the wave equation reduces to the Laplace equation. Similarly, the temporal term in the pressure expression is negligible when $S \ll 1$ which is implied by the previous condition. The compressible potential, expressed in the transformed coordinates, then satisfies the same Laplace equation as the incompressible potential. After the Prandtl–Glauert transformation, the impermeability condition differs from its incompressible counterpart by a multiplicative factor $\gamma_s$. As the Laplace equation is linear, scaling the boundary condition by a factor also scales the corresponding solution by the same factor. Therefore, this leads to $\frac{\partial \phi}{\partial x} = \gamma_s \frac{\partial \phi_{inc}}{\partial x'}$. From a rigorous standpoint, the incompressible solution is associated with the transformed geometry resulting from the Prandtl–Glauert mapping, this corresponds to the Göthert rule. In the classical Prandtl–Glauert approximation, this transformed geometry is replaced by the actual geometry, since the geometric correction is neglected under the thin-body assumption. Hence, the incompressible pressure coefficient is evaluated on the original geometry and the compressible pressure perturbation is therefore equal to $p_{comp} = \gamma_s p_{incomp}$. \newline

As the velocity approaches the characteristic wave speed, this formulation predicts an unbounded increase in pressure due to the divergence of the Lorentz factor, mirroring the behavior encountered in special relativity. In aerodynamics, this effect is commonly referred to as the sound barrier: as an aircraft approaches the speed of sound, pressure levels ahead of the body rise sharply, making further acceleration increasingly difficult. 

\subsection{Validity limit}
Despite this apparent singularity at $M=1$, supersonic flight remains possible. The divergence instead highlights the limits of validity of the present theoretical framework, which relies on assumptions that cease to hold near sonic conditions. In particular, the flow is assumed to remain irrotational and the disturbances induced by the moving body are considered small. These assumptions are violated as nonlinear and rotational effects become significant. For air, this theory generally provides reliable predictions up to Mach numbers of approximately $M\approx 0.6$ for spheres. Some authors therefore developed improved corrections by replacing the Lorentz factor with more sophisticated formula. The Kármán-Tsien \cite{Tsien,Karman} and Laitone \cite{Laitone} rules  give the pressure coefficient $C_p = \frac{P-P_0}{\frac{1}{2} \rho_0 v^2}$ with the following formula

\begin{equation}
C_{p,comp} = \frac{C_{p,incomp}}{\sqrt{1-M^2} + \alpha M^2\frac{C_{p,incomp}}{2}}
\end{equation}

where $\alpha=0$ in the Prandtl-Glauert rule, $\alpha=\frac{1}{1+\sqrt{1-M^2}}$ in the Kármán-Tsien rule, and $\alpha=\frac{1+\frac{\kappa-1}{2}M^2}{\sqrt{1-M^2}}$ in the Laitone rule ($\kappa$ is the heat capacity ratio). These rules allow better predictions up to greater Mach numbers but still fail when shock waves appear. \newline

The primary limitation of the linearised compressible potential flow theory is the onset of shock waves. Even when the freestream velocity is below the speed of sound, the local flow velocity may exceed it at some point on the body surface, giving rise to a local supersonic region terminated by a shock. The freestream Mach number at which the local Mach number first reaches unity on the body surface is called the critical Mach number $M_{cr}$ \cite{Critical_Mach}, and is defined implicitly by the condition \cite{Critical_Mach2}
\begin{equation}
C_{p,cr}(M_{cr}) = C_{p,min}(M_{cr})
\end{equation}
where $C_{p,cr}$ is the pressure coefficient corresponding to $M_{local}=1$, derived from the exact Bernoulli equation, linearisation being invalid at this condition since a sonic point is not a small perturbation of the freestream. $C_{p,min}$ is the minimum pressure coefficient on the body surface in the absence of shocks, estimated from the linearised potential flow theory within its domain of validity, for instance via the Prandtl–Glauert rule. The linearised theory ceases to be applicable for $M > M_{cr}$, beyond which shocks appear and the isentropic and small-perturbation assumptions are simultaneously violated. In such transonic flows, the equations must be modified \cite{potential2}. \newline

The precise computation of the critical Mach number generally requires numerical simulation. However, under the simplifying assumption that the speed of sound remains constant throughout the flow, which is justified for weakly compressible fluids, i.e. fluids with a high speed of sound, an analytical estimate can be derived. Under this assumption, the barotropic relation gives $P-P_0 = c_s^2 (\rho-\rho_0)$. Equation \eqref{eqini2} therefore reads in stationary regime
\begin{equation}
\frac{1}{2}u^2 + \int_{P_0}^{P^*} \frac{d\tilde{P}}{\rho} = C
\end{equation}
Here, $C$ is not equal to 0 as the gauge choice $C=0$ cannot be made in the stationary regime as previously. Because the point considered is where $M=1$, the relation becomes
\begin{equation}
\frac{1}{2}c_s^2 + \int_{P_0}^{P^*} \frac{d\tilde{P}}{\rho} = \frac{1}{2} v^2
\label{bernouil}
\end{equation}
The integral term evaluates to
\begin{equation}
\int_{P_0}^{P^*} \frac{d\tilde{P}}{\rho} = \int_{P_0}^{P^*} c_s^2\frac{d\tilde{P}}{\rho_0 c_s^2 + \tilde{P}-P_0} = c_s^2\ln \left(1+\frac{P^*-P_0}{\rho_0 c_s^2}\right)
\end{equation}
Injecting this relation in \eqref{bernouil} leads to
\begin{equation}
\frac{1}{2} + \ln \left(1+\frac{P^*-P_0}{\rho_0 c_s^2}\right) = \frac{1}{2}\frac{v^2}{c_s^2} = \frac{1}{2} M^2 \Rightarrow C_{p,cr}(M) = \frac{P^*-P_0}{\frac{1}{2}\rho_0 v^2} = \frac{2}{M^2}\left(e^\frac{M^2-1}{2} - 1\right)
\end{equation}
The critical Mach number is then given by the following relation
\begin{equation}
\frac{2}{M_{cr}^2}\left(e^\frac{M_{cr}^2-1}{2} - 1\right) =  \frac{C_{p,incomp,min}}{\sqrt{1-M_{cr}^2}}
\end{equation}
The incompressible minimum pressure coefficient depends solely on the body geometry. For a sphere, the incompressible minimum pressure coefficient is $-\frac{5}{4}$, leading to a critical Mach number of approximately 0.6. For an ellipse of semi-axes $a$ (streamwise) and $b$ (transverse), the surface velocity in incompressible potential flow \cite{Ellipse} is $v\frac{a+b}{\sqrt{a^2\sin^2\theta + b^2\cos^2\theta}}\sin\theta$. The incompressible Bernoulli's principle gives $C_{p,incomp} = 1-\frac{u^2}{v^2} = 1-\left(\frac{a+b}{\sqrt{a^2\sin^2\theta + b^2\cos^2\theta}}\sin\theta\right)^2$. It attains its minimum at $\theta=\frac{\pi}{2}$, which leads to $C_{p,incomp,min} = -\varepsilon^2-2\varepsilon$ with $\varepsilon = \frac{b}{a}$. For a circle, $M_{cr} \sim 0.45$. When $\varepsilon = 0.1$, $M_{cr} \sim 0.84$, and for $\varepsilon = 0.01$, $M_{cr} \sim 0.97$. This general trend holds independently of the fluid properties and reflects a fundamental geometric principle: the thinner the body, the weaker the flow acceleration it induces, and the closer to the speed of sound the linearised compressible potential flow theory remains valid ($\lim\limits_{\varepsilon \rightarrow 0} M_{cr} = 1$). This geometric limit is particularly significant in the context of the analogy with special relativity where the Lorentz transformation remains valid at least up to $v \approx 0.999999991 c$. The body must therefore be vanishingly thin. \newline

Furthermore, for thin bodies, the theory becomes applicable again in the supersonic regime: in this case, the factor $\frac{1}{\sqrt{1-\frac{v^2}{c_s^2}}}$ is replaced by $\frac{1}{\sqrt{\frac{v^2}{c_s^2}-1}}$, except in regions where shock waves form, for which the present assumptions are no longer valid. This is the so-called Ackeret theory \cite{Ackeret,Supersonic,Supersonic2}, mainly developed for airfoils.

\subsection{Doppler effect}
Consider a point source emitting a longitudinal sound wave at frequency $f_0 = \frac{\omega_0}{2\pi}$ in a fluid at rest. As stated in Section 3.1.4, if the source moves at velocity $v$ in the direction $\mathbf{e}_x$, the frequency received by a stationary observer at angle $\theta$ relative to the direction of motion is given by the classical Doppler formula:
\begin{equation}
f = \frac{f_0}{1-M\cos\theta}
\end{equation}
Taking into account compressibility effects, the convected wave equation reduces to the standard wave equation in the transformed coordinate system $(x',y',z',t')$. The emitted waves therefore take the form $cos(\omega_0 t' - \mathbf{k}\cdot \bm{x'})$ where $\lvert \mathbf{k} \rvert = \frac{\omega_0}{c_s}$. The phase of the wave is therefore
\begin{equation}
\Phi = \omega_0 t' - \mathbf{k}\cdot \bm{x'} = \frac{\omega_0}{\gamma_s} t - \omega_0\frac{v\gamma_s}{c_s^2}x - k_x \gamma_s x - k_y y - k_z z
\end{equation}

Therefore, the compressibily of the fluid modifies the oscillating term and adds a factor $\frac{1}{\gamma_s}$ into the frequency \cite{frequency}. In direct analogy with special relativity, the Doppler effect is therefore given by 
\begin{equation} 
f = \frac{f_0}{\gamma_s \left(1-M\cos\theta\right)} 
\end{equation}
which is similar to equation \eqref{Relativistic_Doppler}. Consequently, the Doppler effect in an irrotational flow of a compressible inviscid fluid must also be corrected by the same factor $\gamma_s$, mirroring the relativistic Doppler effect.

\section{Extension of the analogy with special relativity}
The connection between the Prandtl–Glauert and Lorentz transformations has been recognised within the fluid mechanics community for some time, yet it has received little attention from theoretical physicists, which may account for the scarcity of works exploring this analogy beyond its most immediate implications. %Among the few exceptions, Chang-Wei \cite{Chang} investigates parallels between relativistic effects and the compressibility of a hypothetical aether, and Ungs \cite{Ungs} extends the analogy to velocity addition and coordinate transformations. 
The present review goes further: building on \cite{Moi2}, we establish that the mathematical equivalence between special relativity and compressible potential flow theory actually encompasses the entire theoretical framework of both disciplines. A striking consequence of this equivalence is that, within this framework, a relativistic particle and a body immersed in a compressible potential flow are mathematically indistinguishable, the equations governing their dynamics being identical up to the substitution of the speed of light by the speed of sound.

\subsection{Particle dynamics}
Any body moving within a fluid experiences a resistive force. This force generally cannot be computed analytically, except in very simple configurations. It can nevertheless be estimated using the concept of added mass \cite{AddedMass}. \newline

Physically, fluid resistance arises from the mass of the fluid that must be displaced by the moving body. As a first step, let us assume that the fluid is incompressible and initially at rest. An accelerating body of velocity $\bm{v}$ along one direction imparts kinetic energy to the surrounding fluid, given by
$K = \frac{1}{2}\rho \int_V \mathbf{u} \cdot \mathbf{u} \mathrm{d}V = \frac{1}{2} \rho \left( \int_V \frac{\mathbf{u} \cdot \mathbf{u}}{\bm{v} \cdot \bm{v}} \mathrm{d}V \right) \bm{v} \cdot \bm{v} = \frac{1}{2} \rho I \bm{v} \cdot \bm{v}$. In a potential flow, where the velocity potential $\phi$ satisfies Laplace’s equation $\nabla^2 \phi = 0$, the fluid velocity field can be expressed as $u(\mathbf{x},t) = \mathbf{v}(t)\boldsymbol{\nabla}\varphi(\mathbf{x})$. Consequently, the quantity $I$ is an invariant, remaining constant over time.\newline

The rate of change of this kinetic energy is equal to the mechanical power associated with the force representing fluid resistance:
$\frac{\mathrm{d}K}{\mathrm{d}t} = \rho I \bm{v} \cdot \frac{\mathrm{d}\bm{v}}{\mathrm{d}t} = \mathbf{F} \cdot \bm{v}$. The resulting added-mass force can therefore be written as $\mathbf{F_a} = \rho I \frac{\mathrm{d}\bm{v}}{\mathrm{d}t} = m_a \frac{\mathrm{d}\bm{v}}{\mathrm{d}t}$,
where $m_a$ denotes the virtual added mass. This added mass depends only on the geometry of the particle and on the fluid properties, and is therefore constant in time. It is worth noting that the added mass is, in general, a second-order tensor. However, for bodies exhibiting symmetry in the plane perpendicular to the direction of motion, it reduces to a scalar quantity. For the sake of simplicity, the latter is considered in the following. \newline

When viscous forces are negligible, the motion of a body immersed in an incompressible fluid initially at rest is governed by the following equation of motion:

\begin{equation}
\frac{\mathrm{d} (m + m_a)\bm{v}}{\mathrm{d}t} = \mathbf{F} .
\end{equation}

A body of mass $m$ moving in such a fluid therefore behaves identically to a body of mass $m + m_a$ evolving in vacuum. From the observer’s perspective, the quantity $m + m_a$ corresponds to the apparent mass of the body. \newline

As fluid resistance essentially arises from the pressure forces exerted by the fluid on the surface of the body in the direction of motion (assumed to be $\mathbf{e_x}$), it can also be expressed as $\int p \mathbf{dS}\cdot \mathbf{e_x} = m_a \frac{\mathrm{d}\bm{v}\cdot\mathbf{e_x}}{\mathrm{d}t}$.
As demonstrated in Section 3.3, the pressure field in a compressible fluid is equal to the pressure field in the corresponding incompressible fluid multiplied by the factor $\gamma_s$. Since $\gamma_s$ is constant in space, it can be taken outside the surface integral, yielding
\begin{equation}
\int p_{comp} \mathbf{dS}\cdot \mathbf{e_x} = \int \gamma_s p_{incomp} \mathbf{dS}\cdot \mathbf{e_x} = \gamma_s m_a \frac{\mathrm{d}\bm{v}\cdot\mathbf{e_x}}{\mathrm{d}t}
\label{eq_ma_compressible}
\end{equation}
Relation \eqref{eq_ma_compressible} extends the classical definition of added mass to compressible fluids, showing that the corresponding added mass is simply given by $\gamma_s m_a$. This results in an equation of motion for a particle that appears as follows:

\begin{equation}
\frac{\mathrm{d} (m+\gamma_s m_a) \bm{v}}{\mathrm{d}t} = \mathbf{F}
\end{equation}

Let us now assume that the intrinsic mass of the particle $m$ is negligible compared to the added mass $m_a$, a situation that arises when the fluid density is much larger than that of the particle. Under this assumption, the particle dynamics is governed by equation \eqref{eqmotionRR}:
\begin{equation}
\frac{\mathrm{d} \gamma_s m_a \bm{v}}{\mathrm{d}t} = \mathbf{F}
\label{eqmotionRR}
\end{equation}

This expression is formally identical to the relativistic equation of motion in special relativity \eqref{EqmotionRR}, with the particle mass replaced by the added mass.

\subsection{Energy}
Based on the work-energy theorem applied to the equation of motion \eqref{eqmotionRR}, the total energy of the system is deduced like in the special relativity theory (see §3.1.3):

\begin{equation}
E = \gamma_s m_a c_s^2
\end{equation}

This does not correspond to the kinetic energy of the subsonic particle itself, which is negligible by assumption ($m\ll m_a$), but rather the energy needed to move the particle into the fluid. For low velocities, where compressible effects are small, \begin{equation} E\sim m_a c_s^2 + \frac{1}{2} m_a v^2 \end{equation}

The second term $\frac{1}{2} m_a v^2$ represents the kinetic energy of the virtual added mass of fluid, which is the incompressible contribution, while the first term $m_a c_s^2$ accounts for the energy required to compress the fluid during motion. \newline

In this context, while the mass-energy equivalence takes a mathematically identical form to that of special relativity, it acquires another physical meaning: the added mass serves as a convenient representation of the energy imparted to the surrounding fluid.

\subsection{Comparative summary}
Before concluding with a synthesis of all the analogous results, it is worth recalling that the speed of light is given by a relation formally similar to that defining the speed of material waves, \textit{i.e}, $c_{wave}\propto\sqrt{\frac{P}{\rho}}$ where $P$ denotes a quantity homogeneous to a pressure and $\rho$ a density. In particular, it has been shown in \cite{Moi,Moi2} that:

\begin{equation}
c\sim\sqrt{10\frac{(\varepsilon_0 e^{-2})^3(k_BT_{CMB})^4}{\rho_c}}\sim 3\times 10^8~\textrm{m}\cdot\textrm{s}^{-1}
\label{c}
\end{equation}

where $\varepsilon_0$ is the vacuum permittivity, $e$ is the elementary charge, $k_B$ is the Boltzmann constant, $T_{CMB}$ is the temperature of the CMB (Cosmic Microwave Background) and $\rho_c$ is the critical density of the universe \cite{Hubble18}. The parameter $(\varepsilon_0 e^{-2})^3(k_BT_{CMB})^4$ is homogeneous to a pressure in $Pa$. While the mathematical parallel is exact, the two theories differ fundamentally in their physical underpinning: special relativity describes the propagation of light in vacuum, whereas the compressible potential flow framework describes the propagation of acoustic perturbations in a material medium. \newline

Table 1 summarizes all the identified similarities between special relativity and linearised compressible potential flows.

\begin{center}
\setlength\extrarowheight{10pt}
\begin{tabular}{|c|c|c|}
\hline
 & Special relativity & Compressible potential flow \\
\hline
Physical phenomenon & Nearing the speed of the light wave & Nearing the speed of the sound wave \\
\hline
Physical interpretation & Time dilation \& Length contraction & Fluid compressibility \\
\hline
Wave speed & $c \propto \sqrt{\frac{P_{vacuum}}{\rho_{vacuum}}}$ & $c_s \propto \sqrt{\frac{P_{fluid}}{\rho_{fluid}}}$ \\
\hline
Mass & Particle mass $m$ & Apparent particle mass $m_a$ \\
\hline
Lorentz factor & $\gamma = \left(1-\frac{v^2}{c^2}\right)^{-\frac{1}{2}}$ & $\gamma_s = \left(1-\frac{v^2}{c_s^2}\right)^{-\frac{1}{2}}$ \\
\hline
Momentum & $\gamma m \bm{v}$ & $\gamma_s m_a \bm{v}$ \\
\hline
Energy & $\gamma m c^2$ & $\gamma_s m_a c_s^2$ \\
\hline
Doppler effect & $f = f_0 \sqrt{\frac{1+\frac{v}{c}}{1-\frac{v}{c}}}$ & $f = f_0 \sqrt{\frac{1+\frac{v}{c_s}}{1-\frac{v}{c_s}}}$ \\
\hline
Beyond the speed barrier & Not defined & $\gamma_s = \left(\frac{v^2}{c_s^2}-1\right)^{-\frac{1}{2}}$ \\
\hline
\end{tabular}
\captionof{table}{Comparison between special relativity and linearised compressible potential flow.}
\end{center}

\section{Conclusion}
This review has presented a unified and pedagogical description of the mathematical equivalence between the linearised compressible potential flow theory and special relativity. By tracing the parallel historical developments that gave rise to these two frameworks, we have shown that two of the most consequential theoretical constructions of physics share a common mathematical skeleton, despite having developed in near-complete intellectual isolation from one another. \newline

The formal equivalence is embodied by the Prandtl–Glauert–Lorentz transformation, which maps the convected wave equation governing acoustic propagation in a uniform mean flow onto the classical wave equation in a quiescent medium, with the speed of sound playing the role of the speed of light, and the Prandtl-Glauert factor $\left(1-M^2\right)^{-1/2}$ that of the Lorentz factor. We have shown that this isomorphism is not limited to the wave equation alone, but extends to the entire theoretical framework of special relativity: the Lorentz transformation, the relativistic Doppler effect, the fundamental equation of dynamics, and the mass-energy equivalence all find exact mathematical counterparts in the compressible potential flow theory. This equivalence is derived by introducing the concept of added mass and relies on the assumption that the density of the surrounding fluid is large compared to that of the immersed particle. A direct and striking consequence is that a relativistic particle and a body immersed in a compressible potential flow are mathematically indistinguishable within this framework, their governing equations being identical up to the substitution of the speed of light by the speed of sound. The relativistic interpretation is replaced by a purely physical one: the effects traditionally attributed to relativity emerge here as a direct consequence of fluid compressibility. \newline

A natural question then arises: is this exact mathematical parallel a mere coincidence, or does it reflect a deeper physical truth? One may even wonder what course history might have taken had the Michelson–Morley experiment been conducted after the advent of aeronautics. Might the compressible flow analogy have offered an alternative path to the Lorentz transformation? Regardless, we find it fitting to close this review with the second Rule of Reasoning in Philosophy of Isaac Newton: "Therefore to the same natural effects we must, as far as possible, assign the same causes."

%\section*{Conflicts of interest}

%The author declares no competing financial interest.

%\section*{Dedication}

%The manuscript was written through contributions of all authors. All
%authors have given approval to the final version of the manuscript.

% The next command determines the bibliography style. Please do not
% change this.
%\bibliographystyle{unsrtDOI}%crunsrt}
%\bibliographystyle{crunsrt}
%\bibliographystyle{plainnat}
\bibliographystyle{unsrt}
%This calls all references from the .bib
%\nocite{*}

%  This inserts the bib file
\bibliography{Biblio}

\begin{thebibliography}{10}

\bibitem{Michelson}
A.~Michelson and E.~Morley.
\newblock On the {R}elative {M}otion of the {E}arth and the {L}uminiferous
  {E}ther.
\newblock {\em American Journal of Science}, 34(203):333--345, 1887.

\bibitem{Test_RR}
S.~Coleman and S.L. Glashow.
\newblock High-energy tests of {L}orentz invariance.
\newblock {\em Physical Review D}, 59(11), 1999.

\bibitem{Test_RR2}
P.~Delva, J.~Lodewyck, S.~Bilicki, E.~Bookjans, G.~Vallet, R.~Le~Targat, P.-E.
  Pottie, C.~Guerlin, F.~Meynadier, C.~Le~Poncin-Lafitte, O.~Lopez,
  A.~Amy-Klein, W.-K. Lee, N.~Quintin, C.~Lisdat, A.~Al-Masoudi, S.~D\"orscher,
  C.~Grebing, G.~Grosche, A.~Kuhl, S.~Raupach, U.~Sterr, I.~R. Hill, R.~Hobson,
  W.~Bowden, J.~Kronj\"ager, G.~Marra, A.~Rolland, F.~N. Baynes, H.~S.
  Margolis, and P.~Gill.
\newblock Test of special relativity using a fiber network of optical clocks.
\newblock {\em Physical Review Letters}, 118(22):221102, 2017.

\bibitem{Test_RR3}
G.~Saathoff, S.~Karpuk, U.~Eisenbarth, G.~Huber, S.~Krohn, R.~Mu\~noz Horta,
  S.~Reinhardt, D.~Schwalm, A.~Wolf, and G.~Gwinner.
\newblock Improved test of time dilation in special relativity.
\newblock {\em Physical Review Letters}, 91(19):190403, 2003.

\bibitem{Glauert}
H.~Glauert.
\newblock The effect of compressibility on the lift of an aerofoil.
\newblock {\em Proceedings of the royal society A}, 118:113--119, 1928.

\bibitem{Prandtl}
L.~Prandtl.
\newblock General considerations on the flow of compressible fluids.
\newblock {\em NACA Tech. Memo}, 805, 1936.

\bibitem{Kussner2}
H.~G. Küssner.
\newblock General airfoil theory.
\newblock {\em NACA Tech. Memo}, 979, 1941.

\bibitem{n1}
F.~Q. Hu, M.~E. Pizzo, and D.~M. Nark.
\newblock On the use of a {P}randtl-{G}lauert-{L}orentz transformation for
  acoustic scattering by rigid bodies with a uniform flow.
\newblock {\em Journal of Sound and Vibration}, 443:198--211, 2019.

\bibitem{Alembert}
Jean le~Rond~d'Alembert.
\newblock {\em Essai d'une Nouvelle Theorie de la Resistance des Fluides (Essay
  on a New Theory on the Resistance of Fluids)}.
\newblock David l'ainé, 1752.

\bibitem{Euler}
L.~Euler.
\newblock Principles of the motion of fluids.
\newblock {\em Physica D: Nonlinear Phenomena}, 237:1840--1854, 2008.

\bibitem{Lagrange}
J.-L. Lagrange.
\newblock {\em Mémoire sur la {T}héorie du {M}ouvement des {F}luides
  ({M}emoir on the {T}heory of {F}luid {M}otion)}.
\newblock Gauthier-Villars, 1781.

\bibitem{Navier}
C.~L. M.~H. Navier.
\newblock Sur les lois des mouvements des fluides, en ayant égard à
  l'adhésion des molécules ({O}n the laws of motion of fluids taking into
  consideration the adhesion of the molecules).
\newblock {\em Annales de chimie et de physique}, 19:244--260, 1822.

\bibitem{Stokes}
G.~Stokes.
\newblock On the {T}heories of the {I}nternal {F}riction of {F}luids in
  {M}otion and of the {E}quilibrium and {M}otion of {E}lastic {S}olids.
\newblock {\em Transactions of the Cambridge Philosophical Society},
  8:287--319, 1845.

\bibitem{Darrigol}
O.~Darrigol.
\newblock Between {H}ydrodynamics and {E}lasticity {T}heory: {T}he {F}irst
  {F}ive {B}irths of the {N}avier-{S}tokes {E}quation.
\newblock {\em Archive for History of Exact Sciences}, 56:95--150, 2002.

\bibitem{Riemann}
B.~Riemann.
\newblock Über die {F}ortpflanzung ebener {L}uftwellen von endlicher
  {S}chwingungsweite ({O}n the {P}ropagation of {P}lane {A}ir {W}aves of
  {F}inite {W}avelength).
\newblock {\em Proceedings of the Göttingen Academy of Sciences}, 8:43--65,
  1860.

\bibitem{Rankine}
W.~J.~M. Rankine.
\newblock On the thermodynamic theory of waves of finite longitudinal
  disturbances.
\newblock {\em Philosophical Transactions of the Royal Society of London},
  160:277--288, 1870.

\bibitem{Miller}
A.~I. Miller.
\newblock {\em Albert Einstein's Special Theory of Relativity: Emergence (1905)
  and Early Interpretation (1905–1911)}.
\newblock Addison-Wesley, 1981.

\bibitem{Maxwell}
J.~C. Maxwell.
\newblock A dynamical theory of the electromagnetic field.
\newblock {\em Philosophical Transactions of the Royal Society of London},
  155:459--512, 1864.

\bibitem{Lorentz}
H.~A. Lorentz.
\newblock {\em Versuch Einer Theorie der Electrischen und Optischen
  Erscheinungen in Bewegten K{\"o}rpern (Attempt of a Theory of Electrical and
  Optical Phenomena in Moving Bodies)}.
\newblock E. J. BRILL., Leiden, 1895.

\bibitem{Poincare}
H.~Poincaré.
\newblock Sur la dynamique de l'électron ({O}n the dynamics of the electron).
\newblock {\em Rendiconti del Circolo Matematico di Palermo}, 21:129--176,
  1906.

\bibitem{Einstein}
A.~Einstein.
\newblock Zur elektrodynamik bewegter körper ({O}n the electrodynamics of
  moving bodies).
\newblock {\em Annalen der Physik}, 17:891--921, 1905.

\bibitem{Whittaker}
E.~Whittaker.
\newblock {\em A {H}istory of the {T}heories of {A}ether and {E}lectricity (2nd
  ed.)}.
\newblock Thomas Nelson, 1951.

\bibitem{Acuna}
P.~Acuña.
\newblock On the empirical equivalence between special relativity and lorentz's
  ether theory.
\newblock {\em Studies in History and Philosophy of Modern Physics},
  46:283--302, 2014.

\bibitem{Lanchester}
F.~W. Lanchester.
\newblock {\em Aerodynamics}.
\newblock Archibald Constable \& Co. LTD., 1907.

\bibitem{Kutta}
W.~M. Kutta.
\newblock Auftriebskräfte in strömenden flüssigkeiten.
\newblock {\em Illustrierte Aeronautische Mitteilunge}, 6(133):133--135, 1902.

\bibitem{Prandtl2}
L.~Prandtl.
\newblock {\em Tragflügeltheorie I. Mitteilung (Theory of Aerodynamics: Part
  I.)}.
\newblock Nachrichten von der Gesellschaft der Wissenschaften zu Göttingen,
  1918.

\bibitem{Prandtl1}
L.~Prandtl.
\newblock Über flüssigkeitsbewegung bei sehr kleiner reibung.
\newblock In {\em Verhandlungen des III. Internationalen Mathematiker
  Kongresses, Heidelberg}, pages 485--491, Leipzig, 1904.

\bibitem{Ackeret}
J.~Ackeret.
\newblock Air forces on airfoils moving faster than sound.
\newblock Technical report, NACA TM NO.317, 1925.

\bibitem{Kussner1}
H.~G. Küssner.
\newblock Allgemeine {T}ragflächentheorie.
\newblock {\em Luftfahrtforschung}, 17:370--378, 1940.

\bibitem{Tsien}
H.~S. Tsien.
\newblock Two-{D}imensional {S}ubsonic {F}low of {C}ompressible {F}luids.
\newblock {\em Journal of Aeronautical Sciences}, 6(10):399--407, 1939.

\bibitem{Karman}
T.~von Karman.
\newblock Compressible {E}ffects in {A}erodynamics.
\newblock {\em Journal of Aeronautical Sciences}, 8(9):337--356, 1941.

\bibitem{Laitone}
E.~V. Laitone.
\newblock New compressibility correction for two-dimensional subsonic flow.
\newblock {\em Journal of Aeronautical Sciences}, 18(5):350, 1951.

\bibitem{Liu}
L.~Liu.
\newblock A new similarity law for transonic–supersonic flow.
\newblock {\em Physics of Fluids}, 34(8):081705, 2022.

\bibitem{German}
B.~J. German.
\newblock {\em A Riemannian geometric mapping technique for identifying
  incompressible equivalents to subsonic potential flows}.
\newblock Thesis, Georgia Institute of Technology, 2007.

\bibitem{Jeffrey}
J.~Kirkman and T.~Takahashi.
\newblock {\em Revisiting the Transonic Similarity Rule: Critical Mach Number
  Prediction Using Potential Flow Solutions}.
\newblock 2016.

\bibitem{Chandre}
O.~Chandre-Vila, J.-P. Boin, B.~Barriety, Y.~Nivet, J.~Morlier, and
  N.~Gourdain.
\newblock Fast nonlinear static aeroelasticity method for high-aspect-ratio
  wings at different mach regimes.
\newblock {\em Journal of Aircraft}, 60(4):1017--1037, 2023.

\bibitem{Wang}
Y.~Wang, Y.~Liu, Z.~Zhou, and S.~Wang.
\newblock Finite-wing-analogy formula for compressibility correction to
  pressure coefficient of an underwater vehicle model at low mach number.
\newblock {\em Physics of Fluids}, 35(1):016111, 2023.

\bibitem{Hong}
Z.~Hong, X.~Liu, Y.~Fu, Z.~Wang, A.~Bao, and L.~Chen.
\newblock Numerical investigation on acoustic installation effects of blended
  wing-body aircraft.
\newblock {\em Chinese Journal of Aeronautics}, page 104132, 2026.

\bibitem{Mars}
M.~{Carreño Ruiz} and D.~D'Ambrosio.
\newblock Aerodynamic optimization and analysis of quadrotor blades operating
  in the martian atmosphere.
\newblock {\em Aerospace Science and Technology}, 132:108047, 2023.

\bibitem{Nacelle}
A.~Heidebrecht and D.~G. MacManus.
\newblock Surrogate model of complex non-linear data for preliminary nacelle
  design.
\newblock {\em Aerospace Science and Technology}, 84:399--411, 2019.

\bibitem{frequency}
R.~Amiet and W.~R. Sears.
\newblock The aerodynamic noise of small-perturbation subsonic flows.
\newblock {\em Journal of Fluid Mechanics}, 44(2):227--235, 1970.

\bibitem{Taylor}
K.~Taylor.
\newblock A transformation of the acoustic equation with implications for
  wind-tunnel and low-speed flight tests.
\newblock {\em Proceedings of the Royal Society of London. A. Mathematical and
  Physical Sciences}, 363(1713):271--281, 1978.

\bibitem{Chapman}
C.J. Chapman.
\newblock Similarity variables for sound radiation in a uniform flow.
\newblock {\em Journal of Sound and Vibration}, 233(1):157--164, 2000.

\bibitem{Gregory1}
A.~L. Gregory, S.~Sinayoko, A.~Agarwal, and J.~Lasenby.
\newblock An acoustic space-time and the {L}orentz transformation in
  aeroacoustics.
\newblock {\em International Journal of Aeroacoustics}, 14(7):977--1003, 2015.

\bibitem{Gregory2}
A.~Gregory, A.~Agarwal, J.~Lasenby, and S.~Sinayoko.
\newblock Geometric algebra and an acoustic space-time for propagation in
  non-uniform flow.
\newblock {\em Proceedings of the Royal Society A}, 2015.

\bibitem{Du}
Y.~Du, T.~Ma, J.~Cai, and D.~Yang.
\newblock Unifying the stationary and convective {F}fowcs {W}illiams and
  {H}awkings solutions for aeroacoustic predictions via a {L}orentz
  transformation.
\newblock {\em Aerospace Science and Technology}, 142:108657, 2023.

\bibitem{Balin}
N.~Balin, F.~Casenave, F.~Dubois, E.~Duceau, S.~Duprey, and I.~Terrasse.
\newblock Boundary element and finite element coupling for aeroacoustics
  simulations.
\newblock {\em Journal of Computational Physics}, 294:274--296, 2015.

\bibitem{Barucq}
H.~Barucq, N.~Rouxelin, and S.~Tordeux.
\newblock Low-order {P}randtl-{G}lauert-{L}orentz based {A}bsorbing {B}oundary
  {C}onditions for solving the convected {H}elmholtz equation with
  {D}iscontinuous {G}alerkin methods.
\newblock {\em Journal of Computational Physics}, 468:111450, 2022.

\bibitem{SpecialRelativity}
E.~F. Taylor and J.~A. Wheeler.
\newblock {\em Spacetime Physics - Introduction to Special Relativity}.
\newblock W. H. Freeman and Co., 1992.

\bibitem{Mohring}
W.~Möhring.
\newblock On energy, group velocity and small damping of sound waves in ducts
  with shear flow.
\newblock {\em Journal of Sound and Vibration}, 29:93--101, 1973.

\bibitem{Mohring2}
W.~Möhring.
\newblock On vortex sound at low mach number.
\newblock {\em Journal of Fluid Mechanics}, 85:685 -- 691, 1978.

\bibitem{Obermeier}
F.~Obermeier.
\newblock The influence of solid bodies on low mach number vortex sound.
\newblock {\em Journal of Sound and Vibration}, 72(1):39--49, 1980.

\bibitem{Critical_Mach}
M.~V. Cook.
\newblock Chapter 12 - aerodynamic modelling.
\newblock In {\em Flight Dynamics Principles (Third Edition)}, pages 353--369.
  Butterworth-Heinemann, third edition edition, 2013.

\bibitem{Critical_Mach2}
E.L. Houghton, P.W. Carpenter, Steven~H. Collicott, and Daniel~T. Valentine.
\newblock Chapter 8 - airfoils and wings in compressible flow.
\newblock In {\em Aerodynamics for Engineering Students (Seventh Edition)},
  pages 525--573. Butterworth-Heinemann, seventh edition edition, 2017.

\bibitem{potential2}
L.~D. Landau and E.~M. Lifshitz.
\newblock {\em Fluid Mechanics. Course of Theoretical Physics. Vol. 6}.
\newblock 2013.

\bibitem{Ellipse}
J.~Wangand and D.~D. Joseph.
\newblock Potential flow of a second-order fluid over a sphere or an ellipse.
\newblock {\em Journal of Fluid Mechanics}, 511:201--2015, 2004.

\bibitem{Supersonic}
R.~Tempest and L.~Rosenhead.
\newblock Notes on the linearised equation for the velocity potential of the
  steady supersonic flow of a compressible fluid.
\newblock {\em Proceedings of The London Mathematical Society}, pages 197--212,
  1949.

\bibitem{Supersonic2}
R.~Tempest.
\newblock Some physical interpretations of potentials representing supersonic
  motion of compressible fluids.
\newblock {\em Mathematical Proceedings of the Cambridge Philosophical
  Society}, 45:246 -- 250, 1949.

\bibitem{Moi2}
F.~Salmon.
\newblock Unifying physics theories with a single postulate.
\newblock {\em Physics Open}, 23:100258, 2025.

\bibitem{AddedMass}
C.E. Brennen.
\newblock A review of added mass and fluid inertial forces.
\newblock Technical Report CR82.010, Naval Civil Engineering Laboratory, Port
  Hueneme, California, 1982.

\bibitem{Moi}
F.~Salmon.
\newblock Three {A}nalytical {R}elations {G}iving {T}he {S}peed of {L}ight,
  {T}he {P}lanck {C}onstant and {T}he {F}ine-{S}tructure {C}onstant.
\newblock {\em Journal of Physical Science}, 34(1):21--26, 2023.

\bibitem{Hubble18}
Planck collaboration~et al.
\newblock Planck 2018 results: V{I}. {C}osmological parameters.
\newblock {\em Astronomy and Astrophysics}, 641:A6, 2020.

\end{thebibliography}

\end{document}